\documentclass{article}

\usepackage{PRIMEarxiv}

\usepackage[utf8]{inputenc} 
\usepackage[T1]{fontenc}    
\usepackage{hyperref}       
\usepackage{url}            
\usepackage{booktabs}       
\usepackage{amsfonts}       
\usepackage{nicefrac}       
\usepackage{microtype}      
\usepackage{lipsum}
\usepackage{fancyhdr}       
\usepackage{graphicx}       
\usepackage{xcolor}
\usepackage{algorithm}
\usepackage{algpseudocode}
\usepackage{subcaption}
\usepackage{booktabs}
\usepackage{graphicx}
\usepackage{array}
\usepackage{enumitem}
\usepackage{calc} 
\usepackage{multirow}
\graphicspath{{media/}}     

\title{Self-regulated Learning Processes in Secondary Education: A Network Analysis of Trace-based Measures
\thanks{\textit{\underline{Citation}}: 
\textbf{Yixin Cheng, Rui Guan, Tongguang Li, Mladen Raković, Xinyu Li, Yizhou Fan, Flora Jin, Yi-Shan Tsai, Dragan Gašević, and Zachari Swiecki. 2025. Self-regulated Learning Processes in Secondary Education: A Network Analysis of Trace-based Measures. In LAK25: The 15th International Learning Analytics and Knowledge Conference (LAK 2025), March 03–07, 2025, Dublin, Ireland. ACM, New York, NY, USA, 12 pages. https://doi.org/10.1145/3706468.3706502}} 
}

\author{
  Yixin Cheng \\
  Monash University\\
  \texttt{yixin.cheng@monash.edu} \\
   \And
  Rui Guan \\
  Monash University\\
  \texttt{rui.guan@monash.edu} \\
  \AND
  Tongguang Li \\
  Monash University\\
  \texttt{guanliang.chen@monash.edu} \\
  \And
  Mladen Raković \\
  Monash University\\
  \texttt{mladen.rakovic@monash.edu} \\
  \And
  Xinyu Li \\
  Monash University\\
  \texttt{xinyu.li1@monash.edu} \\
  \And
  Yizhou Fan \\
  Peking University\\
  \texttt{fyz@pku.edu.cn} \\
  \And
  Flora Jin \\
  Monash University\\
  \texttt{flora.jin@monash.edu} \\
  \And
  Yi-Shan Tsai \\
  Monash University\\
  \texttt{yi-shan.tsai@monash.edu} \\
  \And
  Dragan Ga\v{s}evi\'c \\
  Monash University\\
  \texttt{dragan.gasevic@monash.edu} \\
  \And
  Zachari Swiecki \\
  Monash University\\
  \texttt{zach.swiecki@monash.edu} \\
}

\begin{document}
\maketitle

\begin{abstract}
While the capacity to self-regulate has been found to be crucial for secondary school students, prior studies often rely on self-report surveys and think-aloud protocols that present notable limitations in capturing self-regulated learning (SRL) processes. This study advances the understanding of SRL in secondary education by using trace data to examine SRL processes during multi-source writing tasks, with higher education participants included for comparison. We collected fine-grained trace data from 66 secondary school students and 59 university students working on the same writing tasks within a shared SRL-oriented learning environment. The data were labelled using Bannert's validated SRL coding scheme to reflect specific SRL processes, and we examined the relationship between these processes, essay performance, and educational levels. Using epistemic network analysis (ENA) to model and visualise the interconnected SRL processes in Bannert's coding scheme, we found that: (a) secondary school students predominantly engaged in three SRL processes---Orientation, Re-reading, and Elaboration/Organisation; (b) high-performing secondary students engaged more in Re-reading, while low-performing students showed more Orientation process; and (c) higher education students exhibited more diverse SRL processes such as Monitoring and Evaluation than their secondary education counterparts, who heavily relied on following task instructions and rubrics to guide their writing. These findings highlight the necessity of designing scaffolding tools and developing teacher training programs to enhance awareness and development of SRL skills for secondary school learners.
\end{abstract}

\keywords{Self-regulated learning \and K-12 education, Epistemic network analysis \and Secondary education}

\maketitle 

\section{Introduction}
An important component of being a proficient learner is the capacity to self-regulate \cite{zimmerman2013}. Zimmerman and Schunk \cite{zimmerman2011} define self-regulated learning (SRL) as a cyclical, constructive, and active process encompassing cognitive, metacognitive, motivational, emotional and behavioural dimensions. SRL has been recognised as an essential aspect of academic success and personal development in secondary school education \cite{dignath2008, kesuma2021}. It is also a crucial factor in understanding barriers during the transition from secondary to higher education \cite{vosniadou2020a}. For example, Vosniadou \cite{vosniadou2020} reported that a considerable proportion of secondary school students lack the adequate knowledge and skills to effectively manage their learning process and prepare themselves as self-regulating learners for the transition to higher education, which implies that their SRL abilities are underdeveloped at the secondary level.

Meanwhile, online and hybrid teaching and learning continues to be prevalent. Research suggests that students who effectively practice SRL strategies are more likely to achieve academic success in online environments \cite{broadbent2015}. Computer-based scaffolds have the potential to offer personalised and adaptable support to aid SRL behaviours \cite{van2023}. For instance, several empirical studies have been conducted that make use of learning analytics to inform assessment \cite{cheng2024}, learning pathways \cite{lim2020}, and feedback \cite{cavalcanti2021}, in support of SRL development. The findings of these studies suggest that collecting trace data---logs containing digital traces of learners' interactions within the learning environment, such as navigational logs, keystrokes, and mouse movements---can infer valid and interpretable constructs of cognitive and metacognitive states. This approach offers a viable means of gaining insights into learning patterns and optimizing SRL strategies for young learners.

Despite significant theoretical and technical advances, there is a notable scarcity of studies using trace data to systematically measure SRL processes in secondary education contexts. This study aims to address this gap by measuring SRL processes among secondary school students engaged in a multi-source writing task using a customised digital writing platform. We also measured SRL processes of higher education students engaged in the same task and used these measures to compare SRL processing between secondary and higher education students. This research hence contributes to a deeper understanding of how secondary school students regulate their learning during writing tasks. The findings may inform targeted educational interventions and teacher education programs aimed at enhancing SRL skills among secondary learners.

\section{Background}

Several conceptual frameworks have been developed to explain SRL \cite{zimmerman2013,bannert2007,winne1998}. These models typically view learners as active agents who regulate and adapt their learning strategies to achieve their goals and respond to different educational contexts \cite{zimmerman2013}. Moreover, these models typically involve a macro-level cyclical processes that include the phases of forethought, performance, and reflection \cite{zimmerman2002becoming}. For example, the COPES model proposed by Winne and Hadwin \cite{winne1998,winne2017} identifies four key phases in the learning process: task definition, goal setting, enacting study tactics and strategies, and adapting studying. Each of these phases is characterised by five learning facets---conditions, operations, products, evaluations, and standards (COPES)---and involves various cognitive and metacognitive processes, such as monitoring, searching, and evaluation. 

Similarly, Bannert \cite{bannert2007} introduced a framework that subdivides SRL into cognitive, metacognitive, and motivational processes. Bannert's framework has been especially useful in examining trace data (e.g., keystroke logging, navigational data) from computer-based environments to gain insights into learners' SRL behaviours \cite{fan2022,fan2022a,li2023,lim2024students,li2024analytics,fan2023towards}. In line with the COPES model, researchers \cite{fan2022} have operationalised the cognitive and metacognitive processes described in Bannert’s framework as sequences of learning actions, including orientation---orienting to the task, planning---determining learning strategies, monitoring---tracking the learning process, and evaluation---self-assessing learning progress---as well as three cognitive processes: first-reading, re-reading, and elaboration/organisation (e.g. writing).

Understanding and supporting the SRL skills of students is important because these skills are related to academic performance \cite{schunk2017,wischgoll2016}. For instance, Karlen \cite{karlen2017} found that high-achieving university students used metacognitive strategies, such as monitoring their writing to meet assignment requirements and revising drafts for content improvement. Similarly, Qin and Zhang \cite{qin2019} discovered that high-performing undergraduates monitored task conditions (e.g., time management and organisation) and revised their compositions during the writing process.

In the context of secondary school, van der Stel and Veenman \cite{van2010} conducted a two-year longitudinal study to investigate the development of metacognitive skills
in history and math problem-solving tasks. The study found that metacognitive skilfulness contributed to learning performance, independent of intellectual ability.
More specifically, they found that the metacognitive skills of young and inexperienced students represent both a general and a domain-specific component. Metacognitive skills predominantly appear to be general with domain-specific component playing a minor role. In terms of general component, the high- and low-performing students were not significantly different. 
In a subsequent longitudinal study, they \cite{van2014} further explored the development of metacognitive skills in adolescents aged 12–15, revealing that these skills do not develop linearly or at the same pace. 
Similarly, Weil et al. \cite{weil2013} found that metacognitive skills—particularly the relationship between task performance and confidence—improves significantly during adolescence, peaking in late adolescence and plateauing into adulthood. These findings suggest that adolescence is a key period for the development of metacognitive skills, which play a vital role in academic performance.


While there have been many studies focusing on the metacognitive skills \cite{van2010,van2014,weil2013,karlen2017}, most  were based on the use of self-report instruments, such as surveys (e.g. the Motivated Strategies for Learning Questionnaire \cite{pintrich1991manual}), and think aloud protocols \cite{bannert2007}. For instance, in \cite{van2010}, students were required to verbalise their thoughts while undertaking two tasks (text studying and problem solving) in two domains (history and math). However, studies (e.g. \cite{veenman2007}) have shown that self-report surveys and think aloud protocols are not suitable for revealing actual SRL processes. Such instruments can trigger process that would not happen otherwise, participants may have limited awareness of what is relevant to report in think alouds, and they may have incomplete or biased recollections of events \cite{barkaoui2011think,bowles2005reactivity}. As noted by Winne and colleagues \cite{winne2017,zhou2012}, trace data can overcome these limitations. 

Previous studies have used trace data to analyse SRL processes \cite{fan2022a,maldonado2018mining}. For example, Fan and colleagues \cite{fan2022}, building on Bannert's theoretical SRL model, developed an iterative approach that combines both theory-driven and data-driven methods to identify SRL processes from trace data. They began by converting the trace data into relevant learning actions---such as Open\_Planner when learners opened a planner tool. Then they used a theory-driven pattern library to interpret the sequences of these actions and map them to SRL processes. A trace parser was used to input the event logs and generate SRL process logs, which were compared with think-aloud data to identify discrepancies and refine invalid interpretations in the pattern library. This iterative process continued until no further significant improvements were observed, resulting in a validated pattern library.

This pattern library has been proven particularly useful for exploring areas such as scaffolding \cite{lim2024students,lim2023effects} and learning strategies \cite{li2024analytics,srivastava2022}. 
For example, Srivastava and colleagues \cite{srivastava2022} conducted a study to identify the SRL processes of higher education students in an academic writing course. Based on writing behaviour patterns analysed via trace data, students were divided into three groups: "Read First, Write Next," "Read and Write Simultaneously," and "Write Intensively, Read Selectively". They found that the first two strategies achieved higher outcomes compared to the third. Specifically, students in the "Read First, Write Next" group focused on cognitive processing of source materials, writing their essays, and monitoring their progress. The "Read and Write Simultaneously" group focused more on comprehending source content and integrating it into their writing products. In contrast, the "Write Intensively, Read Selectively" group primarily focused on writing, dedicating less time to reading, and achieved the lowest outcomes. Regarding SRL processes, the authors found that the "Read First, Write Next" group exhibited equally prevalent elaboration/organisation and first-reading, with more orientation and monitoring than the other two groups. The "Read and Write Simultaneously" group showed a higher prevalence of first-reading, followed by elaboration/organisation. The "Write Intensively, Read Selectively" group predominantly focused on elaboration/organisation in their SRL processes. 

While studies linking academic performance and SRL using trace data have been useful, the majority focus on higher education. Consequently, the relationship between SRL skills and academic performance in secondary education contexts is less understood. 
Findings from higher education may not always apply to secondary education due to differences in cognitive and metacognitive strategies between secondary school students and university students. For instance, in terms of cognitive processes, Coertjens et al. \cite{coertjens2017} found that memory and analysis skills significantly increased after students transitioned from secondary to higher education. In terms of metacognitive processes, Lawanto et al. \cite{lawanto2013} found that college students demonstrated more planning, strategy choice, and monitoring during complex projects compared to secondary students, indicating a more developed use of metacognitive strategies. In Van der Stel and Veenman’s longitudinal study \cite{van2014}, metacognitive skills initially exhibited domain-specific traits but became predominantly general over time, suggesting that adolescence is a critical period in their development.


This gap leaves us with limited understanding of the relationship between SRL and performance in secondary education. To address this gap, we conducted what is to our knowledge the first trace-based study of SRL and performance in a secondary education context. We then compared our results to those obtained from a similar study in a higher education context. Specifically, we asked the following research questions (RQs):

\begin{description}[leftmargin=32pt, labelindent=10pt, style=multiline]
    \item[RQ1] What SRL processes are prevalent, and to what extent do these processes differ between high- and low-performing students in secondary education?
    \item[RQ2] What SRL processes are prevalent, and to what extent do these processes differ between high- and low-performing students in higher education?
    \item[RQ3] To what extent do SRL processes differ across educational contexts?
\end{description}

\section{Method}

\subsection{Experiment Design}
In 2023, we conducted two studies that used the same writing task on the same platform to compare SRL processes between participants in higher education (HE) and secondary education (SE). For the first study, we recruited 59 university participants (36 females, 23 males) from a Chinese public university to explore SRL during a 2-hour essay writing task. The participants (M = 22.63, SD = 3.32) included 28 undergraduates and 31 graduates from 40 majors across 10 disciplines. English was their second language. For the second study, we recruited 66 students (28 females, 37 males, and 1 non-binary) from two secondary schools in Australia to examine SRL among 12-15-year-olds (M = 13.44, SD = 0.84) during a 45-minute multi-source essay task. Thirty-three participants were recruited from each school. All students spoke English as their primary language. Consent forms were signed by teachers, participants, and their parents, and ethical approval was obtained from Monash University.

In the HE study (hereafter referred to as the HE condition), participants were asked to compose a 200-400 word essay about the future of education in 2035, integrating three topics: (1) AI (eight subsections, M = 271 words, SD = 135.97); (2) differentiation in education (three subsections, M = 403 words, SD = 26.17); and (3) scaffolding in education (five subsections, M = 336 words, SD = 77.61). Each set of texts provided relevant information on AI, instructional adaptation, and external support in education.

In the SE study (hereafter referred to as the SE condition), students were asked to write a 200-300 word essay on AI's potential impact on future education. To accommodate secondary students' knowledge levels and cognitive demands, which differ from those of higher students \cite{guerra2017}, we reduced the writing session time to 45 minutes (compared to 2 hours/120 minutes for the HE condition) and simplified the reading materials and rubrics for better accessibility. To ensure the task was suitable, we conducted a pilot study in which a secondary school student reviewed both the materials and the rubrics. In the full study, students used a closed set of nine English texts (M = 178.55 words, SD = 125.24), covering two topics: (1) AI (four subsections, M = 182 words, SD = 166.55), and (2) the school of the future (five subsections, M = 175 words, SD = 102.67), describing how the use of AI-based technologies could improve education.

\begin{figure*}[ht]
  \centering
  \includegraphics[width=\linewidth]{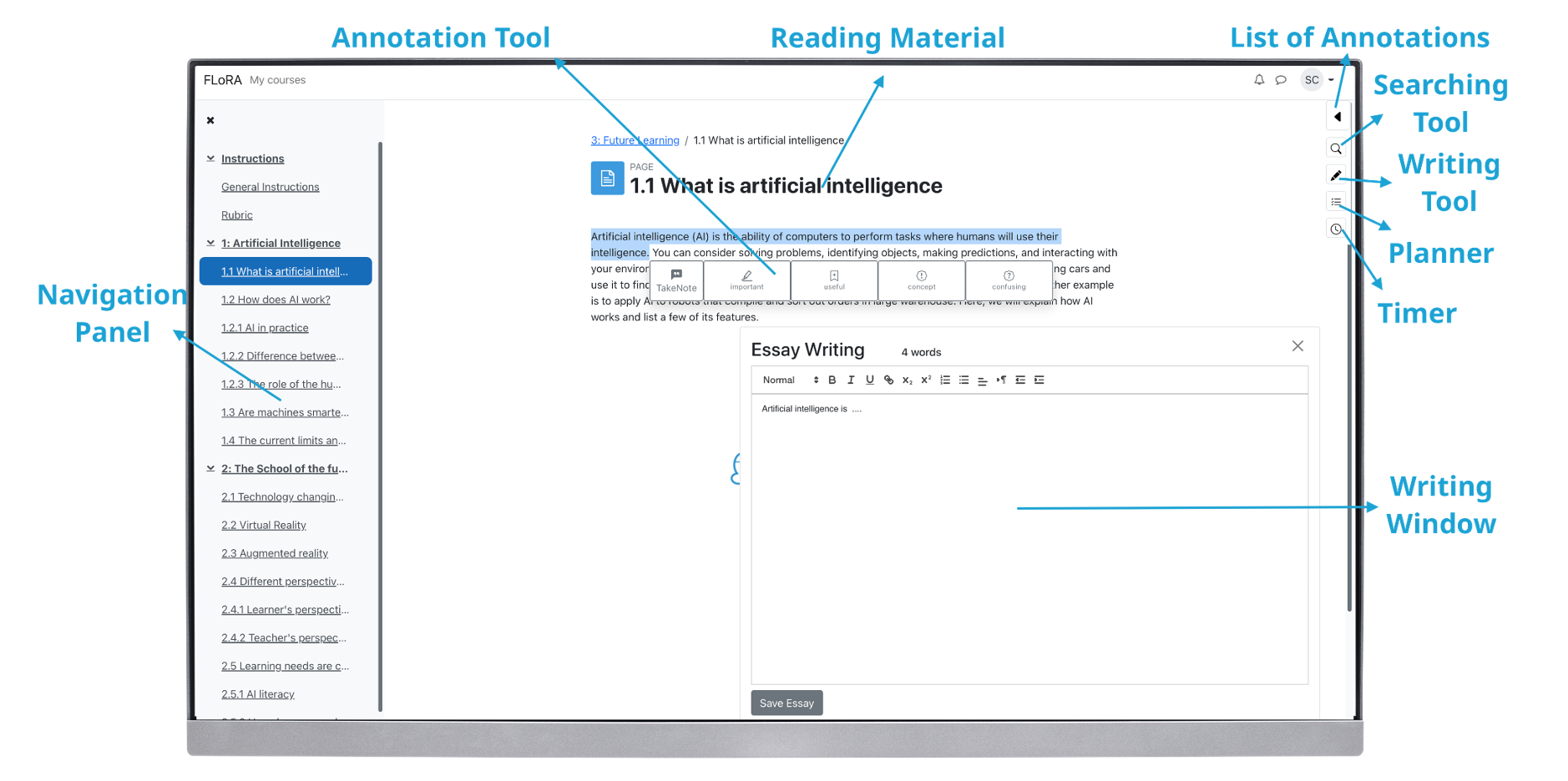}
  \caption{Platform for multi-source writing task}
  \label{cella-platform}
\end{figure*}

To facilitate writing tasks and collect trace data, the participants used a Moodle-based learning platform FLoRA \cite{rakovic2022}, shown in Figure \ref{cella-platform}, specifically developed to support SRL in reading and writing tasks. The platform's layout included a panel for navigating reference texts on the left; instructions, rubrics, and reading materials in the middle; a set of SRL tools (known as instrumentation tools in the literature \cite{van2021}) on the right including a list of annotations made by the participants on the reference texts, a search tool, a writing tool, a planner, and a timer. Participants could compose their essays using a text editor positioned at the bottom-right of the screen.

Prior to the experiment, participants completed a survey to gather demographic information and took a pre-test to assess their prior knowledge of the topics. Following the survey, a training session introduced the interface, task, instructions, and tools to familiarise students with the experimental environment. 

\subsection{Data}

\subsubsection{Survey}
We designed a survey to collect demographic information, such as age and gender, as well as majors and English proficiency measured by the College English Test (CET-4) \cite{rencet4}. The CET-4 test was not included for SE participants as their English proficiency was already assured by their educational background.

\subsubsection{Pretest scores}
We administered a pretest to assess prior knowledge on writing topics before the session for both conditions. For SE condition, the test included 15 multiple-choice questions covering AI, virtual reality, algorithms, and educational concepts like collaborative learning. For HE condition, the test followed the same format but covered recall, understanding, application, and evaluation of concepts in artificial intelligence, differentiation, and scaffolding. Each test was worth a total of 15 points, with each question carrying one point, applying to both conditions.

\subsubsection{Essay scores}
After the participants completed their essays, we collected and scored them manually. For the SE condition, essays were scored out of 18 points using a six-part criterion: AI concept (definition, explanation, examples), AI's role in education (current and future applications), and word count. Each part was rated on a 0-3 scale (no mention, copy, paraphrase, novel information), while word count was scored separately based on the total number of words. Two researchers independently scored 10\% of the essays to assess reliability, targeting a Cohen's \emph{kappa} of 0.8 for each criterion. After two rounds of sampling, all marking criteria met the threshold. Subsequently, the remaining 52 essays were divided between the two researchers for scoring, with participants identified as high or low performers based on the median score (mdn = 11.5).

For the HE condition, each essay was manually graded out of 25 points based on an eight-part scoring scheme: coverage of three topics (1-3), topic integration (4), word count (5), writing skills (6), vision (7), and originality (8). Each part was scored with a maximum of 4 points, except for certain parts, such as word count, which had a maximum score of 3. A random sample of 12 essays was double-scored to assess inter-rater reliability, achieving an intraclass correlation coefficient (ICC) \cite{koo2016} above 0.85. The remaining essays were scored by one researcher, and participants were classified as high or low performers based on the median score (mdn = 14).

\subsubsection{Trace data}
Participant interactions with the platform were automatically recorded as timestamped trace events, which included navigational logs, keystrokes, and mouse movements. In the SE condition (RQ1), a total of 547,812 events were recorded, averaging 8,300 events per session (i.e. per participant). The mean time spent on each session was 38.4 minutes (SD = 8.39). In the HE condition (RQ2), there were 1,329,882 events in total, with an average of 22,540 events per session. The mean time spent on each session was 115.4 minutes (SD = 9.68). For RQ3, data from both conditions were combined for analysis.

\subsection{Self-regulated Learning Process Modelling and Labelling}
\begin{figure*}[ht]
  \centering
  \includegraphics[width=\linewidth]{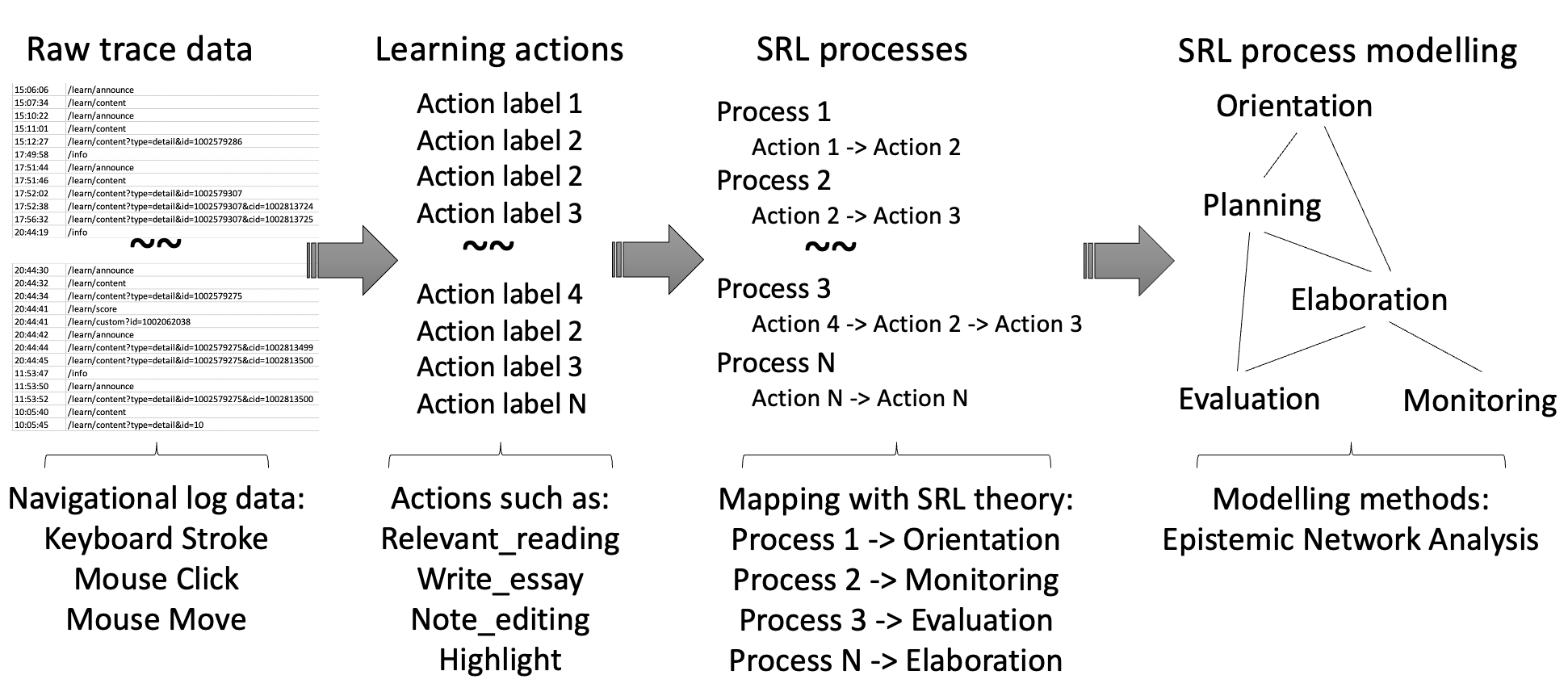}
  \caption{SRL process labelling procedure proposed by \cite{fan2022}}
  \label{fig:srl_mapping}
\end{figure*}

Adapting the approach proposed by Fan and colleagues \cite{fan2022} in Figure \ref{fig:srl_mapping}, we first created a action library that translated the trace events into identifiable actions. In total, we identified seven main actions: ANNOTATION, ESSAY, INSTRUCTION, NAVIGATION, PLANNER, READING, and TIMER. To capture finer-grained dynamics, we defined sub-actions for each main action. For instance, under ANNOTATION, we included sub-actions like Create\_Note and Label\_Annotation. After converting the trace events into actions, we labelled the data to reflect the SRL processes outlined in Bannert’s SRL framework. This coding framework was validated by \cite{fan2022} (see Table \ref{tab:srl_lib}).

To apply the codes to the data we developed a trace parser. This parser identifies 32 sequences of participants actions and maps them to corresponding SRL processes, as delineated in the coding scheme provided in a GitHub repository (\url{https://github.com/yixin-cheng/lak25}). First, a pattern library is constructed using predefined action sequence and the patterns are sorted by the length of their sequences in descending order to prioritise longer sequences. The parser then extracts a subset of actions from the participants' trace data with window size determine by the pattern, then it verifies if the set of actions can be matched to the pattern in the specified order. If a pattern match is found, the process label is assigned to the relevant rows in the trace data for the length of the matched sequence, and the parser continues from the next action outside the sequence, advancing by the length of the matched sequence. If no match is found, the iteration proceeds to the next action. This ensures that all action sequences in the dataset are examined and labelled according to the predefined process patterns. The reliability of the coding scheme is guaranteed as it is based on regular expression matching on a fixed set of patterns that appear in the trace data.

\begin{table}[ht]
\centering
\begin{tabular}{p{2.3cm}p{2cm}p{5.1cm}p{4.2cm}}
\toprule
\textbf{Main Category} & \textbf{SRL process} & \textbf{Definition} & \textbf{Example Pattern} \\
\midrule
\multirow{4}{*}{Metacognition} & \textsc{Orientation} & Orientation on the task and learning activities; Reading of general instructions and rubrics. & After reading the general instruction page, learners read through the navigation to get a overview of what topics they need to learn, then read some pages. \\
\cmidrule{2-4}
 & \textsc{Planning} & Planning of the learning process by arranging activities and determining strategies. Proceeding to the next topic. & Open planner tool and make personal plan. \\
\cmidrule{2-4}
 & \textsc{Evaluation} & Evaluation of the learning process; checking of content-wise correctness (e.g., the essay content) of learning activities. & Check instructions/rubrics when they read their own writing then move on to write or read something else. \\
\cmidrule{2-4}
 & \textsc{Monitoring} & Monitoring and checking the learning process; checking of progress according to instruction or plan. & Check planner or timer; search or read certain annotations; read previous notes during navigation. \\
\midrule
\multirow{2}{*}{Low Cognition} & \textsc{First.Reading} & First time reading information from the materials and superficial describing of pictorial representations. & Read and highlight new learning materials. \\
\cmidrule{2-4}
 & \textsc{Re.Reading} & Rereading of information in the text or figures. & Review relevant learning materials they have read before. \\
\midrule
High Cognition & \textsc{Elaboration. Organisation} & Elaborate by connecting content-related comments and concepts during reading or writing; Organising of content by creating an overview; write down information point by point in notes or essay window; summarising; adding information generated by oneself; and editing information by rephrasing or integrating information with prior knowledge. & Write the essay continuously; Create or choose label for the highlights they made during learning; Write the essay after reading the rubrics. \\
\bottomrule
\end{tabular}
\caption{SRL processes and their definitions with action examples.}
\label{tab:srl_lib}
\end{table}

\subsection{Epistemic Network Analysis}
Given the interconnections between SRL processes proposed in Bannert’s model \cite{saint2022,li2023}, we used ENA \cite{Shaffer2017} to compare the SRL processes of participants in these data. Specifically, we used the \texttt{rENA} package in R \cite{rENA2023}. ENA generates a network representation for each unit of analysis with nodes corresponding to codes and edges representing the frequency of code co-occurrence within the unit's data. The networks were configured as follows:

\begin{itemize}

\item \textbf{Units of Analysis}: Each participant was represented by an individual network, identified by variables for user ID and server ID.
\item \textbf{Codes}: Seven codes were used: \textsc{Monitoring}, \textsc{Orientation}, \textsc{Elaboration.Organisation}, \textsc{First.Reading}, \textsc{Planning}, \textsc{Re.Reading}, and \textsc{Evaluation}, as outlined in Table \ref{tab:srl_lib}.
\item \textbf{Conversations}: Lines (i.e., trace events) were grouped based on variables for user ID and server ID.
\item \textbf{Window Size}: A window size of 50 lines was used in all three models define the boundaries within which codes could co-occur. This window size was determined by manually identifying a sample of 30 events, collected the lengths of the associated windows, and used the median of the distribution to determine a reasonable size.
\end{itemize}

We constructed three ENA models using the means rotation (MR) approach, which identifies the network embedding space whose first dimension maximises between-group variance. The first model compared the means of participants in the high and low performance groups in the SE condition; the second compared the high and low performance groups in the HE condition; and the third compared the means of participants in the different education levels (i.e. SE vs. HE). To interpret the embedding space, the ENA algorithm co-registers the network graphs in the embedding space by aligning the positions of nodes and their connections with the most influential co-occurrences (i.e. prevalence) on each dimension. This allows researchers to visually interpret the dimensions according to the connections at the extremes. For more details, see \cite{Bowman2021}. 

To identify which SRL processes are prevalent in the first two research questions, we examined the participants overall mean network to observe which connections were most prominent---i.e., which had the thickest edges. For all research questions, we compared groups in terms of their mean networks and the mean position of their embeddings---ENA scores---on the first dimension of the corresponding embedding spaces. We used network subtractions to highlight the co-occurrences that were more frequent in one group compared to the other by subtracting the edge wights of mean networks for the two groups.

\subsection{Regression Analysis}

To further address our research questions, we performed regression analyses to statistically compare the ENA scores between the groups of interest while controlling for potential confounds. To address RQ1, we regressed ENA scores for the performance-rotated dimension on a categorical variable for performance and covariates for school, task length, and pretest level (\textit{M1}). To address RQ2, we regressed ENA scores for the performance-rotated dimension on a categorical variable for performance and covariates for school, pretest level, task length, and English language proficiency (CET4) (\textit{M2}). To address RQ3, we regressed ENA scores for the education level-rotated dimension on a categorical predictor for education level and covariates for pretest level and performance (\textit{M3})\footnote{Note that we did not include task length as a covariate in this model, as the time spent in both education levels did not overlap and was thus collinear with the education level predictor.}, as presented in Table \ref{tab:regression_results}. For each model, we tested for significant interactions between the covariates and the categorical variable of interest. Confidence intervals for the regression coefficients were calculated using the percentile method of bootstrapping \cite{Hyndman1996}. Effect sizes for significant differences between the groups of interest were determined using Cohen's $d$ \cite{cohen1988}.

\section{Results}
\subsection{RQ1: SRL Process Difference between SE Conditions}

\begin{figure*}[ht]
  \centering
  \includegraphics[width=\linewidth]{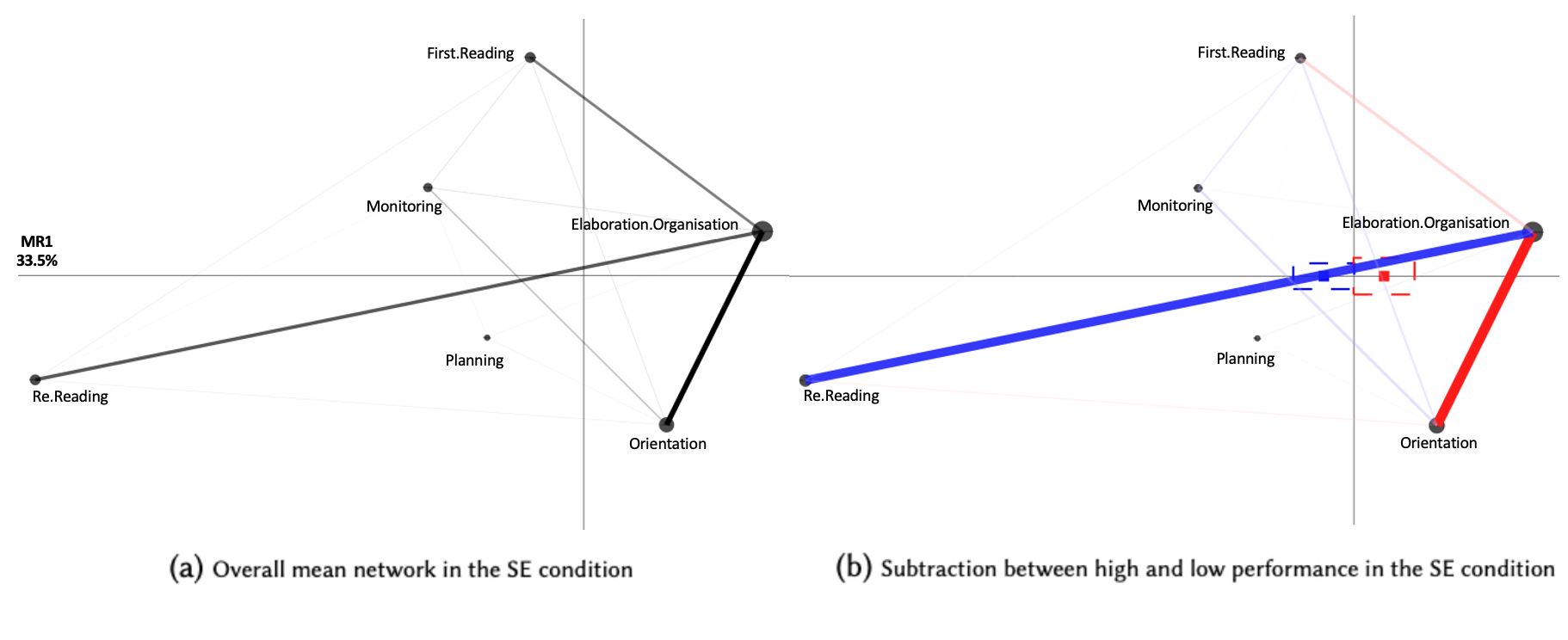}
  \caption{ENA visualisations on SE condition (blue: high performance; red: low performance)}
  \label{fig:ena_se}
\end{figure*}

\begin{table}[htbp]
\centering
\begin{tabular}{p{3cm} >{\centering\arraybackslash}p{2cm} >{\centering\arraybackslash}p{4cm} c}
\hline
 & \textbf{M1 (SE)} & \textbf{M2 (HE)} & \textbf{M3 (SE vs HE)} \\
\hline
Intercept & $\mathbf{-0.49}^{*}$ & $-0.95$ & $\mathbf{-0.41}^{***}$ \\
                   & (0.19)        & (0.88)        & (0.04)           \\
performance\_low & $\mathbf{0.33}^{***}$ & $\mathbf{0.24}^{*}$ & $\mathbf{0.04}^{*}$ \\
                          & (0.10)   & (0.10)      & (0.02)           \\
school\_1 & $0.18$ & -                    & -                    \\
                     & (0.10)        & -                & -                    \\
cet4\_score        & -                & $0.00$ & -                    \\
                     & -                & (0.00)      & -                    \\
level\_SE     & -                & -                & $\mathbf{0.65}^{***}$ \\
                     & -                & -                & (0.02)           \\
pretest\_score       & $0.04$      & $0.01$        & $0.01$             \\
                          & (0.03)   & (0.03)      & (0.00)           \\
task\_length    & 0.00            & -0.00                & -                   \\
                          & (0.00)            & (0.01)                & -                    \\
\hline
Residual & 0.38  & 0.37  & 0.10 \\
R-squared            & 0.19        & 0.20         & 0.91           \\
F-value          & 3.59  & 3.44  & 408.1 \\
p-value              & <0.01      & <0.01        & < 0.001        \\
Num. obs. & 66 & 59 & 125\\
\hline
\multicolumn{4}{l}{\scriptsize{$^{***}p<0.001$; $^{**}p<0.01$; $^{*}p<0.05$}} \\
\end{tabular}
\caption{Regression models with ENA scores on the first dimensions as the dependent variables in performance and level conditions. Standard errors in parentheses. Dashes indicates omitted variables.}
\label{tab:regression_results}
\end{table}

Figure \ref{fig:ena_se} shows the overall mean network and network comparison for the SE condition. The first dimension of the ENA space (MR1\footnote{MR refers to "means rotation", which is a dimension reduction technique that maximise the variance between conditions. In this study, we did not consider the second dimension, as the rotation method ensures the mean value of the conditions is zero in that dimension \cite{Bowman2021}.}) explained 33.5\% of the total variance in the data. This dimension primarily distinguished participants who made connections to \textsc{Re.Reading} on the left versus those who made connections to \textsc{Elaboration.Organisation} on the right. The shift from left to right suggests a process of engaging in writing and note-taking by revisiting previously read information. Notably, connections to \textsc{Evaluation} were not present in SE condition. Further examination of the data revealed that not only were no connections to this code present, but this code never appeared in the SE condition data, indicating that these students did not examine content for correctness during their writing sessions. 

Figure \ref{fig:ena_se}a shows the mean network for both conditions in SE condition. The thickness of the lines between nodes represents the strength—or prevalence—of the connections between these SRL processes. The strongest connections observed were among \textsc{Orientation}, \textsc{Re.Reading}, \textsc{Elaboration.Organisation}. These frequent connections suggest that task understanding, re-reading, and essay writing or note taking are closely linked and prevalent SRL processes for secondary school students during this writing task.

Figure \ref{fig:ena_se}b shows the subtraction between high- and low-performing students. Connections that appeared more frequently in the high- performing students are in blue, while connections that occurred more frequently in low-performing students are in red. The graph shows that high-performing participants tended to revisit relevant learning materials that they have been reading before while writing essay or taking notes, as evidenced by the stronger connections between \textsc{Re.Reading} and \textsc{Elaboration.Organisation}. In contrast, low-performing students were more inclined to focus on incorporating task instructions/rubrics in writing process---suggesting a tendency to use these resources to guide their writing, as indicated by the stronger connection between \textsc{Orientation} and \textsc{Elaboration.Organisation}.

The average ENA scores (blue and red squares)---corroborate the regression analysis (\textit{M1}) in Table \ref{tab:regression_results}, where participants in high-performing condition are significantly farther to the right on MR1 (t= $ 3.38$, $\beta = 0.33$, $p < 0.01$, $d = 0.70$, 95\% CI $[0.17, 0.50]$), indicating that the differences highlighted by the subtraction graph are statistically significant controlling for covariates.

\subsection{RQ2: SRL Process Difference between HE Conditions}

\begin{figure*}[ht]
  \centering
  \includegraphics[width=\linewidth]{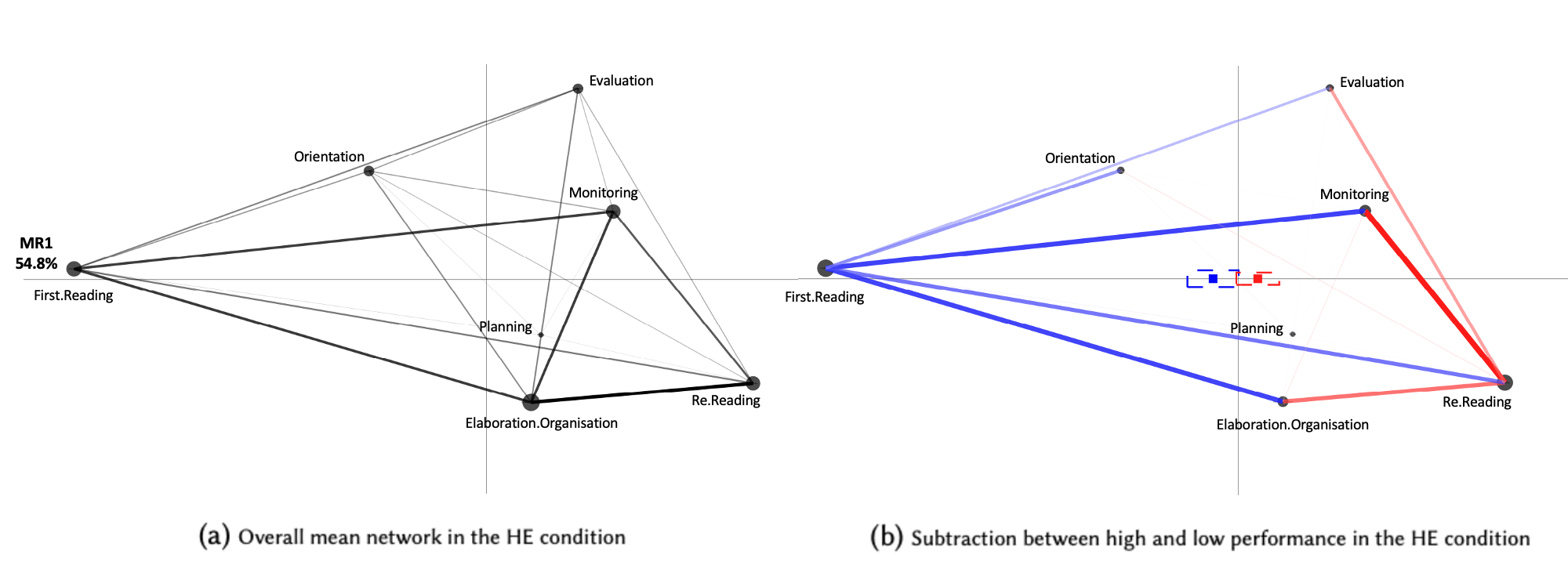}
  \caption{ENA visualisations in HE condition (blue: high performance; red: low performance)}
  \label{fig:ena_he}
\end{figure*}

The ENA embedding space and network comparisons comparing the high and low performance students in the HE condition are shown in Figure \ref{fig:ena_he}. The first dimension of the ENA space explained 54.8\% of the total variance in the data. This dimension distinguishes participants who made connections to \textsc{First.Reading} on the left from those who made connections to \textsc{Re.Reading} on the right. A shift from left to right reflects the sequential reading process---moving from an initial reading to revisiting the given content, suggesting deeper engagement with the material.

Figure \ref{fig:ena_he}a shows the overall mean network for HE group, illustrating a dense network of connections. The varying thickness of the lines suggests the strength of the relationships. Participants in this condition exhibited various prevalent SRL processes, evidenced by the strong connections among \textsc{First.Reading}, \textsc{Elaboration.Organisation}, \textsc{Monitoring}, and \textsc{Re.Reading}. These connections suggest that participants in the HE condition tended to engage in initial reading and re-reading processes, checking timers, and writing essays or taking notes.

Figure \ref{fig:ena_he}b shows the network subtraction for performance groups. Similar to the subtraction graph in SE condition, in this graph, connections that occurred more frequently in the high performance condition are shown in blue, while connections that occurred more frequently in the low performance condition are shown in red. The graph shows that high-performing participants engaged in continuous, page-by-page reading while frequently tracking their progress (e.g., checking timers or reviewing specific annotations), revisiting previous content, regularly incorporating instructions/rubrics into their reading, and actively writing essays or taking notes, evidenced by stronger connections between \textsc{First.Reading} and \textsc{Monitoring}, and among \textsc{Re.Reading}, \textsc{Orientation}, and \textsc{Elaboration.Organisation}. In contrast, participants in the low-performance condition were more inclined to review relevant source materials while checking their own writing against task instructions or rubric, monitoring their reading process by reviewing annotations, and writing essays. This is evidenced by stronger connections between \textsc{Re.Reading} and \textsc{Evaluation}, \textsc{Monitoring}, and \textsc{Elaboration.Organisation}.

The mean ENA scores on MR1 show low performance condition participants are positioned significantly further to the right compared to high performance condition participants, on average. The results of the regression analysis (\textit{M2}) in Table \ref{tab:regression_results} corroborate the visualisation (t= $ 2.31$, $\beta = 0.24$, $p < 0.01$, $d = 0.72$, 95\% CI $[0.07, 0.42]$), indicating a statistically significant difference between the high and low performers controlling for covariates.

\subsection{RQ3: SRL Process Difference between Educational Levels}

\begin{figure*}[ht]
  \centering
  \includegraphics[width=0.5\linewidth]{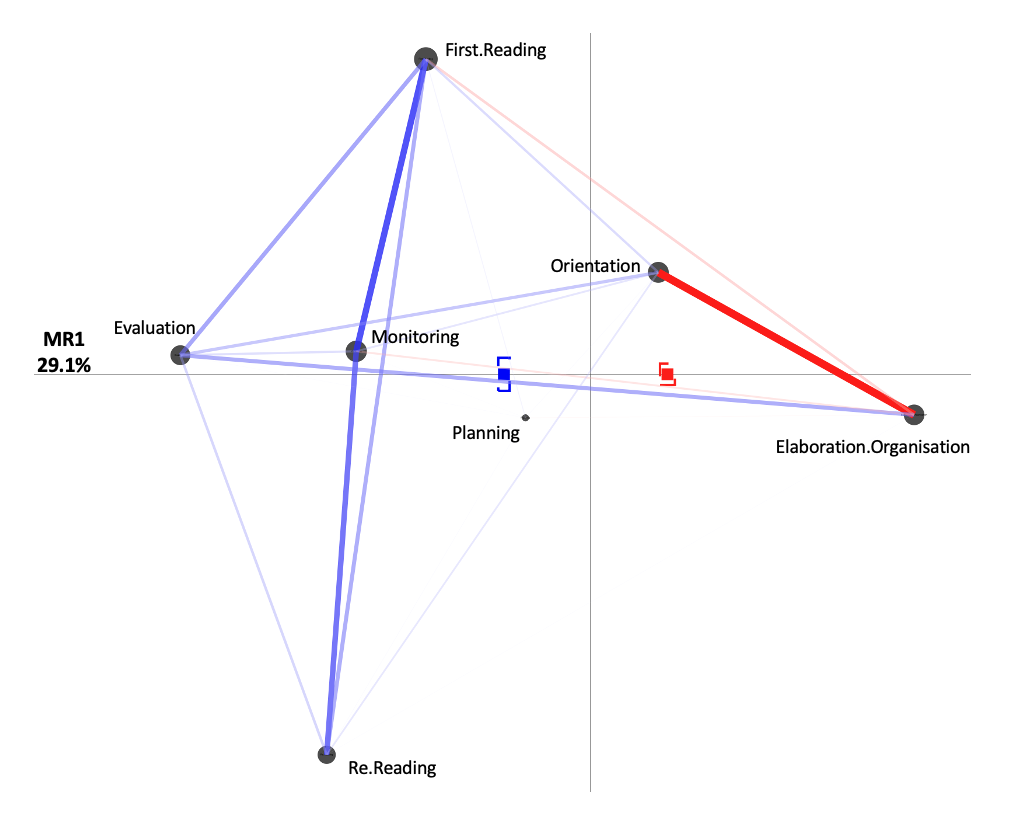}
  \caption{ENA visualisations on subtraction between educational levels (blue: higher education; red: secondary education)}
  \label{fig:ena_all_sub}
\end{figure*}

The ENA embedding space and network comparisons for the HE and SE conditions is shown in Figure \ref{fig:ena_all_sub}. The first dimension explained 29.1\% of the total variance in the data. This dimension distinguishes participants in the HE condition, who made connections to \textsc{Evaluation} and \textsc{Re.Reading} on the left, from those in the SE condition, who made connections to \textsc{Elaboration.Organisation} on the right. This plot also shows the network subtraction for both conditions. Connections that occurred more frequently in the HE condition are shown in blue, while connections that occurred more frequently in the SE condition are in red. The graph shows that participants in the HE condition tended to monitor their reading processes, check the timer, and evaluate their learning processes, as indicated by more frequent connections among \textsc{Monitoring} and \textsc{First.Reading} and \textsc{Re.Reading}; \textsc{First.Reading} and \textsc{Evaluation}. Participants in the SE condition were more inclined towards frequently integrating task instructions/rubrics into their writing process, evidenced by the stronger connection between \textsc{Orientation} and \textsc{Elaboration.Organisation}.

The mean ENA scores on MR1 show that HE condition participants are positioned significantly further to the left compared to SE condition participants, on average. The results of the regression analysis (\textit{M3}) in Table \ref{tab:regression_results} corroborate the visualisation (t = $32.96$, $\beta = 0.65$, $p < 0.001$, $d = 6.14$, 95\% CI $[0.62, 0.68]$), confirming a significant difference between the two conditions controlling for covariates.

\section{Discussion and Conclusions}
In this study, we aimed to measure trace-based SRL processes in secondary education contexts. We conducted two studies: one with 66 secondary school students and another with 59 higher education students, both completing the same multi-source writing task. Using a custom-built platform, we captured detailed trace data, converted it into learning actions, and identified SRL constructs based on Bannert's SRL model \cite{bannert2007}. We then applied ENA and regression analyses to model, visualise, and analyse the SRL processes, addressing the three research questions.



Our first research question sought to identify the prevalent SRL processes among secondary education students and to compare these processes between high- and low-performers. We found that secondary students predominantly engaged in the familiarisation with the task (\textsc{Orientation}), revisited the material to enhance understanding (\textsc{Re.Reading}), and actively processed information through writing and note taking (\textsc{Elaboration.Organisation}). 

We found that the \textsc{Evaluation} process was absent among secondary school students, indicating a potential gap in their ability to self-assess learning process---a critical SRL component as per Bannert's model. Prior studies have identified a positive relationship between self-assessment and learning performance \cite{mcdonald2003impact}, and the effects of self-assessment tools on self-regulation \cite{panadero2012rubrics}, where scripts (e.g., scaffolding questions) were found to be more effective than rubrics for enhancing self-regulation. Although we included rubrics in our study, they did not appear to promote self-assessment in SE as expected. Future studies should consider designing scaffolding tools to support this aspect of SRL in SE.

When comparing high- and low-performing students, we found statistically significant different patterns. High-performing students demonstrated a stronger connection between \textsc{Re.Reading} and \textsc{Elaboration.Organisation}. This implies that they are more likely to engage in iterative processing of information, integrating new knowledge with existing understanding through elaboration. Such behaviour also suggests that students continually refine their comprehension \cite{li2023}. In contrast, low-performing students showed a stronger connection between \textsc{Orientation} and \textsc{Elaboration.Organisation}. This pattern may indicate that these students spent more time understanding task requirements while attempting to process and organise information, possibly due to initial uncertainties regarding the comprehension of the topic or a lack of deep engagement with the material.

Due to the scarcity of trace-based studies in secondary education based on Bannert's model, there are limited insights available for comparison with existing literature. The closest comparison is the study reported in \cite{van2010}, which used think-aloud protocols to extract SRL processes. Contrary to their findings, where general metacognitive skills showed no significant difference between high- and low-performing students, our study revealed significant differences. This discrepancy may be attributed to differences in measurement methods, as think-aloud protocols rely on participants’ verbalisation of their thought processes, which may be incomplete or influenced by their awareness of being observed \cite{barkaoui2011think,bowles2005reactivity}. In contrast, trace data captures fine-grained learning actions without requiring active self-reporting, providing a more objective and continuous measure of SRL behaviours \cite{winne2017,zhou2012}. This allows for the identification of subtle differences in how high- and low-performing students engage with metacognitive strategies that may not be evident through self-reporting or think-aloud methods.


Our second research question focused on identifying the prevalent SRL processes and comparing high and low performance among higher education students. We found strong connections among four dominant SRL processes: \textsc{First-Reading}, \textsc{Re.Reading}, \textsc{Elaboration.Organisation}, and \textsc{Monitoring}, suggesting a richer and more diverse set of SRL strategies compared to their secondary education counterparts. The presence of connections to \textsc{Evaluation} indicates a greater tendency to self-assess their learning progress, likely influenced by rubrics and instructions. In contrast, secondary students may not perceive rubrics as a tool for self-assessment and may lack the foundational skills to benefit from them. Consistent with findings from \cite{van2014}, this finding reinforces developmental differences in metacognitive skills and suggests that, for secondary students, it is necessary to train them on how to use rubrics effectively to evaluate their own work. 

When examining self-regulated learning processes in relation to academic performance in higher education, we found significant differences between high- and low-performing students. High performers centred their activities around \textsc{First.Reading}, \textsc{Elaboration.Organisation}, \textsc{Monitoring}, and \textsc{Orientation}, aligning with the "Read First, Write Next" strategy, which has been linked to higher academic achievement \cite{srivastava2022}. These students focus on fully understanding the materials and rubrics before organising and elaborating on their thoughts in writing, building a strong foundation for composition. In contrast, low performers relied heavily on \textsc{Re.Reading}, suggesting difficulties in comprehending material during initial reading and a need to revisit content multiple times. Their lack of alignment with other strategy groups in \cite{srivastava2022} suggests they may use less structured learning approaches, highlighting a need for further research to better support this subgroup. Furthermore, both high- and low-performing higher education students demonstrated strong \textsc{Monitoring} processes, which contrasts with previous research linking \textsc{Monitoring} to better academic performance \cite{karlen2017,rakovic2022}. This suggests that simply focusing on \textsc{Monitoring} is not enough to differentiate performance levels. High performers likely combined \textsc{Monitoring} with \textsc{First.Reading}, reflecting a more strategic approach by actively checking their understanding during the first read, while low performers paired \textsc{Monitoring} with \textsc{Re.Reading}, indicating that they did not engage deeply enough during the first read and became aware of this deficiency through monitoring. Notably, high performers also exhibited a connection between \textsc{First.Reading} and \textsc{Re.Reading}, indicating that re-reading was a productive process for these learners. During the first reading, high performers revisited previously read content to integrate new knowledge with prior information, confirming and refining their understanding. These findings highlight the importance of understanding how self-regulated learning processes connect rather than viewing them in isolation \cite{saint2022}.


In addressing RQ3, which compared SRL processes between educational levels, the statistically significant differences highlight the developmental progression of SRL skills. Higher education students demonstrated greater engagement in processes including \textsc{Monitoring}, \textsc{Evaluation}, \textsc{First.Reading}, and \textsc{Re.Reading}, indicating stronger metacognitive awareness and an ability to self-regulate their learning more effectively. These students strategically used tools, adjusting their approach based on ongoing assessments of their understanding. This result aligns with research showing that metacognitive skills become more sophisticated with age \cite{van2014} and in more complex educational contexts \cite{lawanto2013}, where students take on greater responsibility for their learning \cite{zimmerman2011}. In contrast, secondary students' focus on \textsc{Orientation} and \textsc{Elaboration.Organisation} suggests they relied on the task instructions and rubrics to understand the task requirements and guide their writing but still need to develop more self-regulation strategies. These findings corroborate and supplement those of Vosniadou \cite{vosniadou2020} by providing nuanced insights into how secondary students do engage with SRL processes but often lack a comprehensive understanding of how these processes interconnect, which limits their effectiveness. This may result from less exposure to tasks requiring self-regulation, more structured learning environments, difficulties in understanding the task requirements, lack of awareness, or a reduced sense of ownership over their learning. Introducing interventions to promote skills such as monitoring and evaluation could help bridge this gap. By fostering these skills earlier, educators can better prepare students for the demands of higher education, where self-regulation is critical for academic success.

Our study has several limitations. First, the difference in task duration between the two educational levels: SE students had 45 minutes, while HE students had two hours. Although our results, as shown in Table \ref{tab:regression_results}, showed that time spent on the task did not significantly affect SRL processes within each condition, this discrepancy complicates comparisons of SRL processes, particularly for RQ3. Longer task duration may have allowed higher education students to engage in processes like monitoring and evaluation, while secondary students may have been constrained by time, limiting their use of such strategies. This suggests that differences in SRL processes might be partly due to the variation in task length rather than educational level alone. Future research should either standardise task duration across groups or control for time (e.g. first or last 10\% of task time) to allow for more accurate comparisons of SRL processes.

Second, the study did not examine SRL processes for SE students at the level of specific learning strategies, as Srivastava and colleagues \cite{srivastava2022} did in the HE context. This highlights the need for future research to explore how SRL processes differ based on various learning strategies.

Third, our study did not examine how SE students transition to HE in their use of SRL processes. It is unclear whether this transition involves a drastic shift or a gradual development of strategies. Future research should investigate when and how this transition occurs to better support students during this critical period.

Fourth, our study focused on group-level comparisons, which may overlook individual differences in SRL processes. Group effects inherently disregard the self-regulated nature of SRL, and statistically significant differences do not always reflect practical relevance. Future research should incorporate qualitative perspectives, such as interviews or observations, to explore individual strategies and their practical implications. Finally, as no prior studies have used trace data to extract SRL processes based on Bannert's model \cite{bannert2007} in secondary education contexts, our ability to compare and contextualise the findings within existing literature is limited.

Despite these limitations, the study identified prevalent SRL processes among secondary school students and significant differences between high- and low-performing students, as well as across educational levels. We hope these findings will inform and inspire future research and practice, particularly regarding the development of SRL skills in secondary students navigating increasingly digital learning environments.

\section*{Acknowledgments}
This research was funded partially by the Australian Government through the Australian Research Council (project number DP240100069 and DP220101209) and the Jacobs Foundation (CELLA 2 CERES). 

\section*{Citation}
Yixin Cheng, Rui Guan, Tongguang Li, Mladen Raković, Xinyu Li, Yizhou Fan, Flora Jin, Yi-Shan Tsai, Dragan Gašević, and Zachari Swiecki. 2025. Self-regulated Learning Processes in Secondary Education: A Network Analysis of Trace-based Measures. In LAK25: The 15th International Learning Analytics and Knowledge Conference (LAK 2025), March 03–07, 2025, Dublin, Ireland. ACM, New York, NY, USA, 12 pages. https://doi.org/10.1145/3706468.3706502

\bibliographystyle{unsrt}  
\bibliography{reference}

\end{document}